\documentclass[12pt]{article}
\usepackage{amssymb,amsmath,amsthm,amsfonts,amscd}
\usepackage{graphicx}
\newcommand\be{\begin{equation}}
\newcommand\ee{\end{equation}}
\newcommand\bea{\begin{eqnarray}}
\newcommand\eea{\end{eqnarray}}

\newcommand{\fatalpha}{{\bf \alpha \kern -0.44em \alpha}}
\newcommand{\fatsigma}{{\bf \sigma \kern -0.54em \sigma}}
\newcommand{\tpchi}{{\bf \chi \kern -0.35em \chi}}
\newcommand{\llambda}{{\bf \lambda \kern -0.45em \lambda}}

\bibliography{plain}
\title{\bf Quantum phase transition in the  Dzyaloshinskii-Moriya interaction
with inhomogeneous magnetic field: Geometric approach}
\author{ G. Najarbashi $^{a}$
 \thanks{Najarbashi@uma.ac.ir} ,B. Seifi $^{a}$ \thanks{B.seifi@uma.ac.ir } \\
\\ \\
$^a${\small Department of Physics, University of Mohaghegh Ardabili, Ardabil 179, Iran.}\\
{\small}} \pagebreak
\begin{document}
\maketitle
\newpage %\vspace{15mm}

\begin{abstract}
\par

%In this paper we have generalized the result of  S. Oh (Physics Letters A. 644-647 \textbf{373 }) to  Dzyaloshinski-Moriya (DM) model under nonuniform external magnetic field  and have investigated its quantum phase transition. To this end, we calculate the concurrence measure and geometric phase (Berry phase) of ground state in to the quaternion representation. In quaternion representation, the two-qubit pure states can be mapped in usual one-qubit states living on Bloch sphere. It has been shown that the geometric phase and  ground state entanglement are complementary systems that can exhibit quantum phase transition in DM model.

In this paper, we  generalize the results of  S. Oh (Physics Letters A. 644-647 \textbf{373 }) to  Dzyaloshinski-Moriya model under nonuniform external magnetic field  to investigate the relation between entanglement, geometric phase (or  Berry phase) and quantum phase transition. We use quaternionic representation to relate the geometric phase to the  quantum phase transition.  For small values of DM parameter, the Berry phase is more appropriate  than the concurrence measure, while   for large values, the concurrence   is a good  indicator to show the phase transition. On the other hand, by  increasing the DM interaction the phase transition occurs for large values of anisotropy parameter. In addition, for small values of magnetic field the concurrence measure is appropriate indicator for quantum phase transition, but for large values of magnetic field the Berry phase shows a sharp changes in the phase transition points. The results show that the Berry phase and concurrence form a complementary system from phase transition point of view.
\par
 {\bf PACs Index: 03.67.a 03.65.Ud}
\end{abstract}
\newpage
\section{Introduction}
Phase transition is a nonanalytic change in the ground state energy as a function of system's parameters is associated with level crossings or avoided crossings between the ground and exited energy levels \cite{Sachdev} and these phase transition points are important in physics. In classical systems there are formal rules to determine phase transition \cite{Goldenfeld}, but in quantum systems this is an open problem. Quantum phase transition is a phase transition in the zero temperature of quantum  systems \cite{Sachdev,Oh}. So it is interesting to find quantum mechanical quantity that can determine the level crossing. In this paper, we will study the quantum phase transition in spin chain system, and we use geometric phase (Berry phase) and concurrence measure to determine this phase transition.
\par
 Single qubit  pure states  can be identified by points on the surface of the  Bloch sphere $S^{2}$, and mixed states are characterized by points inside the Bloch sphere. The generalization of this concept in two and three-qubit states are described by the  Hopf fibration. The relation between Hopf fibration, single qubit and two-qubit states, has  been studied by Mosseri and Dandoloff  \cite{Mosseri}  in quaternionic skew-field and subsequently have been generalized to three-qubit state based on octonions by Bernevig and Chen  \cite{Bernevig} . However, there is also one more reason to look for Hopf fibration and  stereographic projections. For two qubit pure states the concurrence measure appears explicitly in quaternionic stereographic projection  which geometrically means that non-entangled states are mapped from $S^7$ onto a 2-dimensional planar subspace of the target space ${\mathbb{R}}^{4}$. On the other hand, it has been shown that the quaternionic representation has a geometric description of geometric phase. The geometric phase is the magnetic flux due to magnetic monopoles located at the level crossing points \cite{Berry,Shapere}. In quaternionic representation of quantum state \cite{Mosseri,Bernevig,Najarbashi1,Najarbashi2}, Levay provided a  elegant interpretation of the geometric phase as the parallel transformation of quaternionic spinors due to Mannoury-Fubini-Study metric in Hilbert space of two qubit states \cite{Levay,Najarbashi3}.  The relation between geometric phases, phase transition and level crossings for the Heisenberg XY model with transverse magnetic field has been investigated by Oh et al \cite{Oh1}.
\par
The entanglement property is one of the most fascinating features of quantum mechanics and this property provides a fundamental resource in quantum information theory \cite{Bennett1,Bennett2,Ekert,Murao}. The entanglement  has been discussed at the early years of quantum mechanics as a specifical quantum computation and quantum information \cite{Schrodinger,Einstein,Bell,Maleki}. In spin chain systems the entangled subsystems of  whole  vector states cannot be separated into a product of the subsystem states. A measurement on one subsystem in quantum entangled system not only gives information about the other subsystem, but also provides possibility of manipulating it.
 Therefore entanglement becomes the main tool in quantum computations, quantum phase transition, quantum cryptography, information processing, teleportation and etc.\cite{Angelakis}.
\par
The single qubit gates are local operators and it is clear that the local operators unable to generate entanglement in an  N-qubit system. To generate entanglement state in N-qubit system we  need an inter-qubit interaction such as  a two qubit gates. The simplest two qubit interaction is described by the Ising interaction between spin half particles in the form of $J_{z}\sigma_{1}^{z}\sigma_{2}^{z}$. More general interaction between two qubits is given by the Heisenberg  model with magnetic field and Dzyaloshinskii-Moriya (DM)  interactions.
Recently entanglement of two qubits and its dependence on external magnetic fields, anisotropy and temperature have been considered in several Heisenberg models \cite{Gunlycke,Yang,Lagmago,Wang,Sun,Kao,Zhu,Asoudeh,Zhang,Zhang1,Kargarian}.
\par
This paper studies the behavior of  quantum correlations and quantum phase transition in the anisotropic XYZ spin-half chain with  uniform and nonuniform external magnetic field and DM interaction  $(\vec{D}.(\vec{\sigma}_{1}\times\vec{\sigma}_{2}))$ \cite{Dzyaloshinsky,Moriya,Kheirandish}. The DM  interaction arising from extension of the Anderson superexchange interaction theory by including the spin-orbit coupling, it is important  for the weak ferromagnetism and for the spin arrangement in antiferromagnetic of low symmetry. It also plays a significant role in performing universal quantum computation \cite{Wu,Wu1}.
 In this state we find nonanalytic dependence of concurrence measure \cite{Wootters,Hill} and geometric phase on the DM interaction, and establish their relations with the quantum phase transition.
In addition, we will show in some regions that entanglement is not a appropriate indicator for the phase transition, the geometric phase is  a good one, and vice versa. In other words, geometric phase and the ground state entanglement are complementary systems that can exhibit quantum phase transition in spin chain systems.
\section{Heisenberg XYZ model with Dzyaloshinskii-Moriya interaction}
In this section, we study the quantum phase transition in a system of two qubits  Heisenberg XYZ model with Z-component DM coupling and non-uniform external magnetic field.
\subsection{The model}
The Hamiltonian of the system is read as
\begin{equation}\label{hamiltonian}
H =   - \frac{{1 + \gamma }}{2}\sigma _1^x\sigma _2^x  - \frac{{1 - \gamma }}{2}\sigma _1^y\sigma _2^y - J_{z}\sigma _1^z\sigma _2^z -\frac{D_{z}}{2}(\sigma _1^x\sigma _2^y-\sigma _1^y\sigma _2^x)-\frac{(\lambda+b_{z})}{2}\sigma _1^z-\frac{(\lambda-b_{z})}{2}\sigma _2^z,
\end{equation}
where $\gamma $ is an anisotropy factor, $J_{z}$ is a real coupling coefficient, $D_z$ is the Z-component Dzyaloshinskii–Moriya (DM) coupling parameter, $\lambda$  and $b_z$,  are  uniform and nonuniform external Z-component magnetic field parameters respectively, $\sigma^{a}_{i}$ are the Pauli matrices of the i'th qubit with $a= x, y, z$. The coupling constants $J_z > 0$ corresponds to the  ferromagnetic  case, and $J_z < 0$ corresponds to the antiferromagnetic case. The Hamiltonian (\ref{hamiltonian}) is the general form of a Heisenberg  Hamiltonian, which is exactly solvable and becomes a
paradigmatic example in the study of quantum phase transitions. The matrix form of Hamiltonian (\ref{hamiltonian}) can be written as:
\begin{equation}\label{mhamiltonian}
H= \left( {\begin{array}{*{20}{c}}
{ - {\lambda} - {J_z}}&0&0&{ - \gamma}\\
0&{{b_z} + {J_z}}&{ - 1 - i{D_z}}&0\\
0&{ - 1 + i{D_z}}&{ - {b_z} + {J_z}}&0\\
{ - \gamma}&0&0&{{\lambda} - {J_z}}
\end{array}} \right)=H^{even}+H^{add}.
\end{equation}
One may define the Hamiltonian
\begin{equation}\label{evenn}
{H^{even}} = \left( {\begin{array}{*{20}{c}}{ - {\lambda} - {J_z}}&{ - \gamma}\\
{ - \gamma}&{{\lambda} - {J_z}}
\end{array}} \right),
\end{equation}
 on the subspace spanned by $\{|00\rangle, |11\rangle$\}. It is easy to write down the eigenvalues and eigenvectors of $H_{even}$ as
\begin{align}\label{evenspectrum}
\begin{array}{c}
{E_ \pm ^e = -{J_z} \pm \sqrt {\lambda^2  + {{\gamma}^2}} }, \\
{\left| {E_ + ^e} \right\rangle  = \cos (\frac{{{\theta _1}}}{2})\left| {00} \right\rangle  + \sin (\frac{{{\theta _1}}}{2})\left| {11} \right\rangle }, \\
{\left| {E_ - ^e} \right\rangle  = \sin (\frac{{{\theta _1}}}{2})\left| {00} \right\rangle  - \cos (\frac{{{\theta _1}}}{2})\left| {11} \right\rangle },
\end{array}
\end{align}
where $\tan (\theta_{1})=\frac{-\lambda}{\gamma}$. On the other hand, the Hamiltonian $H^{odd}$ on the subspace  $\{|01\rangle, |10\rangle$\} is given by
\begin{equation}\label{odd}
{H^{odd}} = \left( {\begin{array}{*{20}{c}}
{{b_z} + {J_z}}&{ - 1 - i{D_z}}\\
{ -1 + i{D_z}}&{{\lambda} - {J_z}}
\end{array}} \right).
\end{equation}
The spectrum of ${H^{odd}}$ is easily obtained as
\begin{align}\label{oddspectrum}
\begin{array}{c}
{E_ \pm ^o = {J_z} \pm \sqrt {b_z^2 + D_z^2 + {{1}}} }, \\
{\left| {E_ + ^o} \right\rangle  = \cos (\frac{{{\theta _2}}}{2})\left| {01} \right\rangle  + {e^{-i\varphi }}\sin (\frac{{{\theta _2}}}{2})\left| {10} \right\rangle }, \\
{\left| {E_ - ^o} \right\rangle  = \sin (\frac{{{\theta _2}}}{2})\left| {01} \right\rangle  - {e^{-i\varphi }}\cos (\frac{{{\theta _2}}}{2})\left| {10} \right\rangle },
\end{array}
\end{align}
\begin{figure}
  \centering
  \includegraphics[width=16 cm]{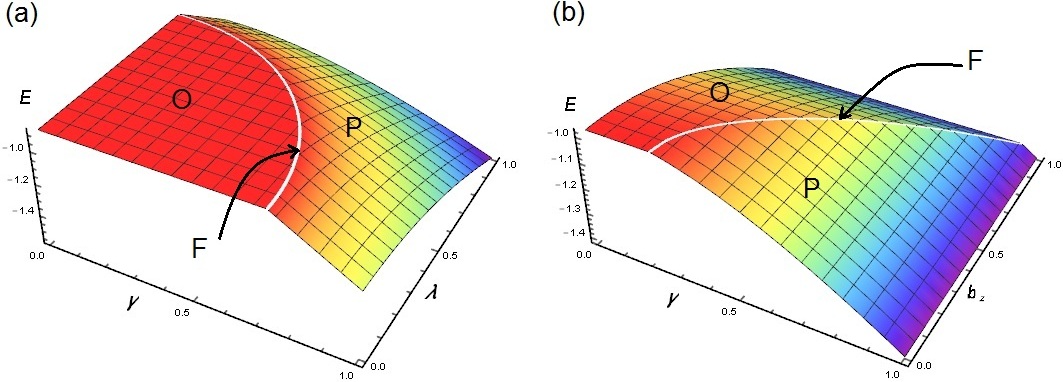}\\
  \caption{\small { (Color online.) (a) Ground energy as a function of $\gamma$ and $\lambda$ for Hamiltonian (\ref{hamiltonian}). The level crossing (white line) for parameters $J_{z}=0.15$, $D_{z}=0.2$, $b_{z}=0.1$.  (b) Ground energy as a function of $\gamma$ and $b_z$ for Hamiltonian (\ref{hamiltonian}). The level crossing (white line) for parameters $J_{z}=0.1$, $D_{z}=0.4$, $\lambda=0.85$. }}\label{appendix Fig.1}
\end{figure}
where $\tan(\theta _2)=\frac{\sqrt {D_z^2 + {{1}}} }{b_{z}}$ and $\tan(\varphi)=D_{z}$. The even and odd eigenvectors of Hamiltonian (\ref{hamiltonian}) confine to the subspace of
even and odd parity operator  $\sigma_{1}^{z}\otimes \sigma_{2}^{z}$, respectively, i.e. $\sigma_{1}^{z}\otimes \sigma_{2}^{z}|E_ \pm ^e\rangle=|E_ \pm ^e\rangle$  and $\sigma_{1}^{z}\otimes \sigma_{2}^{z}|E_ \pm ^o\rangle=-|E_ \pm ^o\rangle$
. Consider the Hamiltonian (\ref{hamiltonian}), whose degrees of freedom reside on the sites of a lattice, and which varies as a function of a dimensionless coupling, $\gamma,J_{z},\lambda,b_{z}$ and $D_{z}$. At zero temperature limit, the system occupies its ground state $E_ - ^o$ or $E_ - ^e$. For the case of a finite lattice, this ground state energy will generically be a smooth and analytic function of Hamiltonian couplings. In $E^{e}_ {-}=E^{o}_ {-}$,  the level crossings occur between the ground and first exited states. An avoided level-crossing between the ground and an excited state of Hamiltonian in a finite lattice could become progressively sharper as the lattice size increases, leading to a non-analyticity at $E^{e}_ {-}=E^{o}_ {-}$ in the infinite lattice limit. We shall identify any point of non-analyticity in the ground state energy of the Hamiltonian system as a quantum phase transition: The non-analyticity could be either the limiting case of an avoided level-crossing or an actual level-crossing. Corresponding to $E^{e}_ {-}<E^{o}_ {-}$, $E^{e}_ {-}=E^{o}_ {-}$ and $E^{e}_ {-}>E^{o}_ {-}$ the system stay at paramagnetic (P), ordered ferromagnetic (F)  and  the oscillatory phase (O), respectively  (see Fig. (\ref{appendix Fig.1})). In Fig. (\ref{appendix Fig.1}-a) the ground state energy is plotted with respect to  $\gamma$ and  $\lambda$, the ordered feromagnetic line shows quantum phase transition points, which by increasing the external magnetic field the phase transition occurs for small value of anisotropy parameter $\gamma$. To show the importance of inhomogeneous magnetic field $b_z$ in quantum phase transition we plot the Fig.(\ref{appendix Fig.1}-b) which  implies that by increasing the inhomogeneous magnetic field the quantum phase transition occurs for large value of anisotropy parameter.
\subsection{Quaternionic representation and Hopf fibration}
 Consider the $\mathcal{H}_{4}^{\mathbb{C}}$ for 4-dimensional  complex Hilbert space which is the tensor product of the individual Hilbert spaces $\mathcal{H}_{2}^{\mathbb{C}}\otimes\mathcal{H}_{2}^{\mathbb{C}}$ with a direct product basis $\{|00\rangle, |01\rangle, |10\rangle, |11\rangle\}$. A general two-qubit pure state reads
\begin{equation}\label{2qubit}
|\psi\rangle=a_{0}|00\rangle+a_{1}|01\rangle+a_{2}|10\rangle+a_{3}|11\rangle,\quad\quad a_{0},...,a_{3}\in{\mathbb{C}}.
\end{equation}
This state is called separable, if it can be represented in the product form
$|\psi\rangle= |\psi\rangle_{A}\otimes|\psi\rangle_{B}$, where $|\psi\rangle_{A}\in\mathcal{H}_{2}^{\mathbb{C}}$ and $|\psi\rangle_{A}\in\mathcal{H}_{2}^{\mathbb{C}}$. This occurs if and only if there exists only one nonzero Schmidt coefficient, $\lambda_{1}=1$, i.e. the reduced state $\rho_{A}$ or $\rho_{B}$  is pure. In the opposite case the state $|\psi\rangle$ is called entangled. The  normalization condition $|a_{0}|^{2} +|a_{1}|^{2} +|a_{2}|^{2}+ |a_{3}|^{2} = 1$ identifies $\mathcal{H}_{4}^{\mathbb{C}}$ to the seven dimensional real
sphere $S^{7}$, embedded in $\mathbb{R}^{8}$. In geometric point of view, the unit sphere $S^{7}$ can be parameterized in many different ways as a product of
manifolds, but for understanding the geometry of two-qubit entanglement it is useful to fibre $S^{7}$ over the four dimensional sphere $S^{4}$ with $S^{3}$ fibres by employing the second Hopf fibration. This idea can be illustrated by introducing a quaternionic representation for two-qubit state, i.e. Using the quaternionic skew-field $\mathbb{Q}$, we can equivalently restate every $|\psi\rangle\in\mathcal{H}_{4}^{\mathbb{C}}$
by a quaterbit  $|\psi\rangle_{\mathbb{Q}}\in\mathcal{H}_{2}^{\mathbb{Q}}$ as
\begin{equation}\label{quaterbit}
|\psi\rangle_{\mathbb{Q}}:=q_{0}|0\rangle_{\mathbb{Q}}+q_{1}|1\rangle_{\mathbb{Q}},
\end{equation}
where $q_{0}=a_{0}+a_{1}\textbf{j}$ and $q_{1}=a_{2}+a_{3}\textbf{j}$ are two quaternion numbers. In quaternion Hilbert space $\mathcal{H}_{2}^{\mathbb{Q}}$ the state  (\ref{2qubit}) can be recast as (\ref{quaterbit}) with the following representation
\begin{align}\label{q-basis}
\begin{array}{c}
|00\rangle\ \ \ \  \longrightarrow \ \ \ \ \ \ |0\rangle_{\mathbb{Q}},\\
|01\rangle\ \ \ \ \ \longrightarrow \ \ \ \ \ \textbf{j}|0\rangle_{\mathbb{Q}},\\
|10\rangle\ \ \ \  \longrightarrow \ \ \ \ \ \ |1\rangle_{\mathbb{Q}},\\
|11\rangle\ \ \ \ \  \longrightarrow \ \ \ \ \   \textbf{j}|1\rangle_{\mathbb{Q}}.
\end{array}
\end{align}
Quaternion is an associative and non-commutative algebra of
rank 4 on real space $\mathbb{R}$ whose every element can be written as $q=q_{0}+q_{1}\textbf{i}+q_{2}\textbf{j}+q_{3}\textbf{k} \in \mathbb{Q}$, where the quaternionic units $\textbf{i},\textbf{j}$ and $\textbf{k}$ with squares equal to -1 satisfy the usual relations $\textbf{i}\textbf{j} = -\textbf{j}\textbf{i} =\textbf{k}$, and similar ones obtained by employing cyclic permutations of the symbols \textbf{i}\textbf{j}\textbf{k}.
 The  quaternion can be equivalently defined in term of complex numbers $z_{1}=q_{0}+q_{1}\textbf{i}$  and  $z_{2}=q_{2}+q_{3}\textbf{i} $ in the form  $q=z_{1}+z_{2}\textbf{j} $. The conjugate quaternion $\bar{q}$ is obtained by $ \bar{q} = (q_{0}- q_{1}\textbf{i})- (q_{2} +q_{3}\textbf{i})\textbf{j}$.
 Note that in term of quaternion numbers the normalization condition of state (\ref{quaterbit}) is given by $|q_{0}|^{2}+|q_{1}|^{2}=1$. Now we are  define the second Hopf fibration by the map as the composition of a stereographic projection $\mathcal{P}$ from $S^7$ to
 $\mathbb{R}^4+\{\infty\}$,
 followed by an inverse stereographic projection $\mathcal{S}$ from $\mathbb{R}^4+\{ \infty \}$ to $S^2$:
\begin{equation}\label{hopf}
\begin{array}{*{20}{c}}
{\begin{array}{*{20}{c}}
{{\cal P}:\quad \quad }&{\quad \quad \left\{ {\begin{array}{*{20}{c}}
{{S^7}\quad \quad \;\quad \longrightarrow {\mathbb{R}^4} + \{ \infty \} }\\
{({q_1},{q_2})\quad \;\; \longrightarrow Q = {q_2}{{\bar q}_2}}
\end{array}} \right.}&{\quad \quad \quad {q_1},{q_2} \in \mathbb{Q} ,}
\end{array}}\\
{\begin{array}{*{20}{c}}
{\quad \quad {\cal S}:\quad \quad }&{\quad \quad \left\{ {\begin{array}{*{20}{c}}
{{{\mathbb{R}}^4} + \{ \infty \} \quad  \longrightarrow {S^4}\quad \quad \quad \quad }\\
{Q\quad \quad \quad  \; \longrightarrow M({x_i})\quad \quad }
\end{array}} \right.}&{\quad \quad \quad \sum\limits_{i = 0}^4 {x_i^2}  = 1 , }
\end{array}}
\end{array}
\end{equation}
or explicitly the fibration  $\mathcal{P}$ maps the state $|\psi\rangle_{\mathbb{Q}}$ as
\begin{equation}\label{p-map}
\mathcal{P}|\psi\rangle_{\mathbb{Q}}:=\frac{q_{0}\bar{q_{1}}\textbf{j}}{|q_{1}|^2}=\frac{{S}+\mathcal{C}\textbf{j}}{|q_{1}|^2},
\end{equation}
where ${S}=a_{0}\bar{a}_{2}+a_{1}\bar{a}_{3}$ and  $\mathcal{C}=a_{0}a_{3}-a_{1}a_{2}$ denote respectively the Schmidt and concurrence terms in quantum information theory. In Hopf fibration (\ref{hopf}) the $x_{i}$ are  Cartesian coordinates for $S^4$ and define as fallow:
\begin{align}\label{s-cor}
\begin{array}{c}
x_{0}=|q_{0}|^{2}-|q_{1}|^{2}\\
x_{1}=2 Re({S}),\\
x_{2}=2 Im({S}),\\
x_{3}=2 Re(\mathcal{C}),\\
x_{4}=2 Im(\mathcal{C}).
\end{array}
\end{align}
 For ${S}=0$  the two-qubit pure state  (\ref{2qubit})  has  Schmidt form  $ \sqrt{\lambda}|00\rangle+\sqrt{1-\lambda}|11\rangle$. On the other hand, for two-qubit pure state   $C=2|\mathcal{C}|$ is concurrence measure and for $C=0$ the two qubit pure state (\ref{2qubit}) is reduced to a separable state, which implies that the    base space is restricted to $S^{2}$ for non-entangled  two-qubit state.
\par
According to Eq. (\ref{q-basis}), the quaternionic form of ground state for$E^{e}_ {-}<E^{o}_ {-}$ and $E^{e}_ {-}>E^{o}_ {-}$ are given by
\begin{equation}\label{q-state}
\left\{ {\begin{array}{*{20}{c}}
{\left| {{E^{e}_{-} }} \right\rangle  \longrightarrow {{\left| {{E^{e}_{-} }} \right\rangle }_{\mathbb{Q}}} = \sin (\frac{\theta_{1} }{2}){{\left| 0 \right\rangle }_{\mathbb{Q}}} - \cos (\frac{\theta_{1} }{2})\textbf{j}{{\left| 1 \right\rangle }_{\mathbb{Q}}},} \quad \quad   &{}\\
{\left| {{E^{o}_{-} }} \right\rangle  \longrightarrow {{\left| {{E^{o}_{-}}} \right\rangle }_{\mathbb{Q}}} =\sin (\frac{\theta_{2} }{2})\textbf{j}{{\left|0\right\rangle }_{\mathbb{Q}}} -e^{-i\varphi}\cos (\frac{\theta_{2} }{2}) {{\left| 1 \right\rangle }_{\mathbb{Q}}}.}&{}
\end{array}} \right.
\end{equation}
According to Hopf fibration (\ref{p-map}),  the concurrence of ground state energy takes the form
\begin{equation}\label{g-state}
C_{(ground\;state)}=\left\{ {\begin{array}{*{20}{c}}
{\sin (\theta_{1} )}&{E^{e}_ {-}<E^{o}_ {-},}\\
{\sin (\theta_{2} )}&{E^{e}_ {-}>E^{o}_ {-}.}
\end{array}} \right.
\end{equation}
\begin{figure}
  \centering
  \includegraphics[width=16 cm]{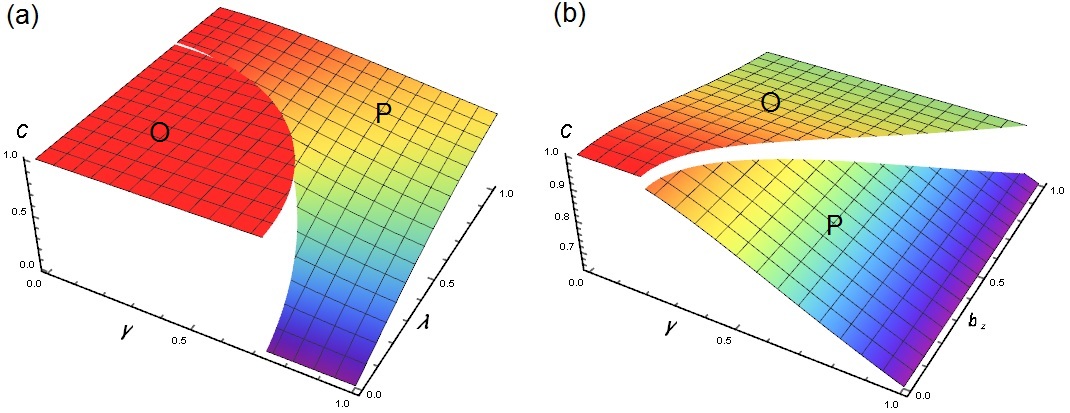}\\
 \caption{\small { (Color online.) (a) Concurrence of ground  state as a function of $\gamma$ and $\lambda$ for Hamiltonian (\ref{hamiltonian}). The level crossing (white line) for parameters $J_{z}=0.15$, $D_{z}=0.2$, $b_{z}=0.1$. (b) Concurrence of ground  state as a function of $\gamma$ and $b_z$ for Hamiltonian (\ref{hamiltonian}). The level crossing (white line) for parameters $J_{z}=0.1$, $D_{z}=0.4$, $\lambda=0.85$ . }}\label{appendix Fig.2}
\end{figure}
As shown in  Fig. (\ref{appendix Fig.2}-a), the entanglement changes abruptly in phase transition point for $\lambda=0$ and $0<\gamma<1$ then it seems that the entanglement works  well as an indicator to quantum phase transitions in this region. But for $\gamma=0$ and $0<\lambda<1$ the concurrence has not tangible  change and it is not a suitable indicator for detect the phase transition. There is similar behavior in  Fig. (\ref{appendix Fig.2}-b), i.e. for small value of inhomogeneous magnetic field $b_{z}$ and $0<\gamma<1$ the concurrence has intangible change, and for large value of $b_{z}$ the concurrence changes abruptly in phase transition point.
\subsection{geometric phase and it's geometrical structure in quantum phase transition}
\subsubsection{geometric phase on one-qubit Bloch sphere}
\begin{figure}
  \centering
  \includegraphics[width=12 cm]{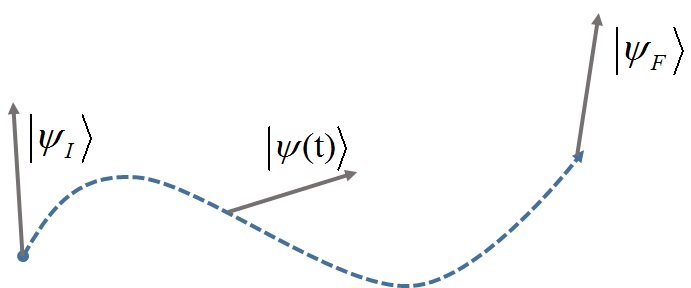}\\
 \caption{\small {(Color online.) Transportation of quantum state on geodesic.}}\label{appendix Fig.transport}
\end{figure}

Quantum states are represented as vectors in a complex vector space, these vectors are only defined up to a global phase which is a unit modulus complex number.  Look at the amplitude between the two states $|\psi_{I}\rangle$ and $|\psi_{F}\rangle$ in the polar  decomposition:
\begin{equation}\label{amplitude}
\left\langle {{{\psi _I}}}
 \mathrel{\left | {\vphantom {{{\psi _I}} {{\psi _F}}}}
 \right. \kern-\nulldelimiterspace}
 {{{\psi _F}}} \right\rangle =r e^{i\xi _{IF}},
\end{equation}
where the $\xi _{IF}$ is the relative phase between the two  states. The states  ${e^{i\alpha _{1}}}\left| {{\psi _I}} \right\rangle $ and ${e^{i\alpha_{2} }}\left| {{\psi _F}} \right\rangle $, which differ from the original states by an  overall arbitrary phase, have a different relative phases by the  amount of $\Delta \alpha  = \alpha_{1}  - \alpha_{2} $.  There are infinitely  choices for $\Delta \alpha$ and they all look equally appropriate which formally says
 that this definition of phase is gauge dependent  (phase dependent).
Consider the path connecting the two states, $|\psi{(t)}\rangle$,  such that when $t=0$ we have $|\psi_{I}\rangle$ and when $t=1$  we have $|\psi_{F}\rangle$. One can transport the states $|\psi{(t)}\rangle$ from the  position $I$ to the position $F$ and see how different the final phase is  to that of $|\psi_{F}\rangle$ through interference. If the states $|\psi_{F}\rangle$ and $|\psi_{I}\rangle$  transported to $|\psi_{F}\rangle$ interfere constructively then they are in phase,  and  the degree  of interference can define the phase difference  (see Fig. (\ref{appendix Fig.transport})) \cite{Vedral}.

From employing the differential geometry,  we know that the transport itself doesn't introduce any  additional "twists and turns" in the phases so that we are actually  comparing some different phases to the original ones. Suppose, we have a curved manifold  and we have a  vector at a point $I$ and another at a point $F$. The relative phase between the two vectors can be measured by  transport one of them to the other one, which the angle between the two vectors is relative phase. The straightest  possible path is known as a geodesic, and the corresponding   evolution along this path is known as the parallel transport. To define  a parallel transport, let's look at the infinitesimal evolution, from $|\psi{(t)}\rangle$ to $|\psi{(t+dt)}\rangle$. If we don't want there to be any twists and  turns in the phase, even infinitesimally, then the two states should  be in phase. So, we require that $Arg[\left\langle {{\psi (t)}}  \mathrel{\left | {\vphantom {{\psi (t)} {\psi (t + dt)}}} \right. \kern-\nulldelimiterspace}  {{\psi (t + dt)}} \right\rangle ]=0$. This is the same as asking that $\left\langle {{\psi (t)}}  \mathrel{\left | {\vphantom {{\psi (t)} {\psi (t + dt)}}} \right. \kern-\nulldelimiterspace}  {{\psi (t + dt)}} \right\rangle $ be purely real, i.e.  up to second order the $Im[\left\langle {{\psi (t)}}  \mathrel{\left | {\vphantom {{\psi (t)} {\psi (t + dt)}}} \right. \kern-\nulldelimiterspace}  {{\psi (t + dt)}} \right\rangle ]= Im[\left\langle {{\psi (t)}}  \mathrel{\left |d| {\vphantom {{\psi (t)} {\psi (t)}}}  \right. \kern-\nulldelimiterspace}  {{\psi (t )}} \right\rangle ] = 0$ . But $\left\langle {{\psi (t)}}  \mathrel{\left |d| {\vphantom {{\psi (t)} {\psi (t)}}}  \right. \kern-\nulldelimiterspace}  {{\psi (t )}} \right\rangle$ is purely imaginary, hence the parallel transport condition becomes $\left\langle {{\psi (t)}}  \mathrel{\left |d| {\vphantom {{\psi (t)} {\psi (t)}}}  \right. \kern-\nulldelimiterspace}  {{\psi (t )}} \right\rangle=0$ . This definition of parallel transport is not automatically gauge invariant. By this  we mean that if instead of the state $|\psi(t)\rangle$, we use the state
\begin{equation}\label{amplitude1}
|\psi'(t)\rangle=e^{i\alpha_{t}}|\psi(t)\rangle,
\end{equation}
 then the parallel transport condition changes by the amount
\begin{equation}\label{amplitude2}
\left\langle {{\psi '(t)}}  \mathrel{\left |d| {\vphantom {{\psi '(t)} {\psi '(t)}}}  \right. \kern-\nulldelimiterspace}  {{\psi '(t )}} \right\rangle=\left\langle {{\psi (t)}}  \mathrel{\left |d| {\vphantom {{\psi (t)} {\psi (t)}}}  \right. \kern-\nulldelimiterspace}  {{\psi (t )}} \right\rangle+i \frac{d\alpha}{dt}dt,
\end{equation}
as can easily be checked. In order to obtain something that is gauge  invariant we can integrate the expression $\left\langle {{\psi (t)}}  \mathrel{\left |d| {\vphantom {{\psi (t)} {\psi (t)}}}  \right. \kern-\nulldelimiterspace}  {{\psi (t)}} \right\rangle$ over a closed loop,  giving us the expression for the geometric phase, and then   exponentiate the result. So, the geometric phase resulting  from the parallel transport is
\begin{equation}\label{berry}
B = \int\limits_i^f {\left\langle {\psi (t)} \right|\frac{d}{{dt}}\left| {\psi (t)} \right\rangle } dt,
\end{equation}
and its exponential (over a closed loop) is gauge independent, but  not path independent.  It is also interesting that the  underlying space is curved and it is the curvature that is reflected in  the phase difference; in fact, the curvature is the phase difference up  to a constant factor. When a quantity vanishes infinitesimally, but  its integral over a finite region does not, then this quantity is called  non-integrable. Therefore, geometric phases are a manifestation of  non-integrable phase factors in quantum mechanics. Let's look at  two level systems  to illustrate this point.
\begin{figure}
  \centering
  \includegraphics[width=9 cm]{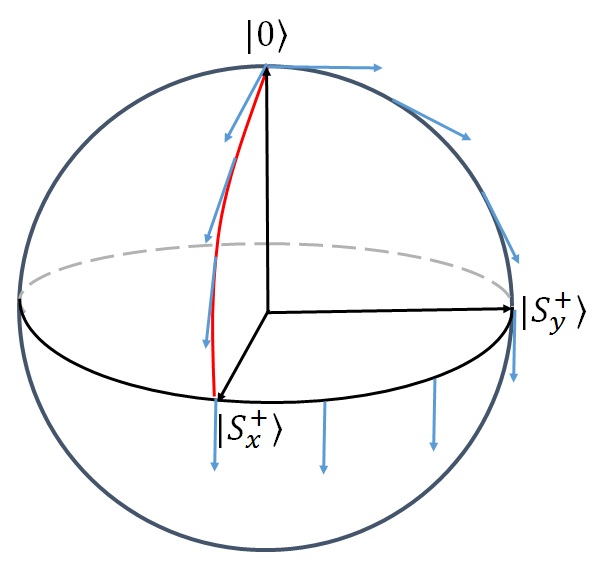}\\
 \caption{\small {(Color online.) Parallel transport of the phase on the Bloch sphere. }}\label{appendix Fig.bloch}
\end{figure}
Suppose that we now evolve from the state $|0\rangle$ to the state  $|S_{x}^{+}\rangle=\frac{1}{\sqrt{2}}(|0\rangle+|1\rangle)$, then to  $|S_{y}^{+}\rangle=\frac{1}{\sqrt{2}}(|0\rangle+i|1\rangle)$ and finally back to the  state $|0\rangle$. On the Bloch sphere, we are going from the north pole to  the equator, then we move on the equator by an angle of $\pi/2$ and  finally we move back to the north pole.  To see  the   corresponding geometric phase we start with a tangential vector initially at the north pole pointing in some direction. If we now parallel transport this vector  along the described path, then we end up with a vector pointing  in a different direction to the original one (even though, infinitesimally, the phase vector has always stayed parallel to itself). The  angle between the two is $\pi/2$, which is exactly equal to the area  covered by the state vector during the transport, or the   corresponding solid angle of the transport (see Fig. (\ref{appendix Fig.bloch})).
It is interesting that for two-qiubit pure states there are Bloch  sphere representation in quaternionic Hilbert space and we can generalized the concept of parallel transform from complex Bloch sphere to quaternionic Bloch sphere.
\par
\subsubsection{geometric phase in two-qubit states and quaternion representation}
For  discussion of geometric phases and criticality in spin-chain systems, we are interested in the Hamiltonian that can be obtained by applying a rotation by angle $\eta$, around the z direction, to each spin, i.e.  $H'=H'^{even}+H'^{odd} = U_z^\dag (\eta )HU_z (\eta)$ where $U_z (\eta)=exp[ - i\frac{\eta}{2}(\sigma _1^z + \sigma _2^z)]$. Then the rotational Hamiltonian is given by
\begin{align}\label{2mhamiltonian}
\begin{array}{c}
H'^{even}=H^{even}=\left(
\begin{array}{cc}
 -2 b_z-j_z & 2 i D_z+1 \\
 -2 i D_z+1 & 2 b_z-j_z \\
\end{array}
\right) , \\
H'^{odd}=\left(
\begin{array}{cc}
 2 \lambda+j_z & \gamma e^{2 i \eta }  \\
\gamma e^{-2 i \eta }  & j_z-2 \lambda \\
\end{array}
\right)\ \ \ \ \ \ \ \ \  \ \ \ \ \ \  \ \   ,
\end{array}
\end{align}
where the $H^{even}$ is invariant under transformation, then the instantaneous ground state $|\psi_{0}\rangle$ satisfying $H'(r)|\psi_{0}\rangle=E_{0}|\psi_{0}\rangle$, and according to (\ref{q-basis}) the transformed ground state reads
\begin{equation}\label{qrstate}
\left\{ {\begin{array}{*{20}{c}}
{ {{\left| {{E'^{e}_{-} }} \right\rangle }_{\mathbb{Q}}} = e^{- i \eta } \sin (\frac{\theta_{1} }{2}){{\left| 0 \right\rangle }_{\mathbb{Q}}} -  e^{ i \eta }\cos (\frac{\theta_{1} }{2})\textbf{j}{{\left| 1 \right\rangle }_{\mathbb{Q}}}} \quad \quad   &{E^{e}_ {-}<E^{o}_ {-}},\\
{ {{\left| {{E'^{o}_{-}}} \right\rangle }_{\mathbb{Q}}} =\sin (\frac{\theta_{2} }{2})\textbf{j}{{\left| 0\right\rangle }_{\mathbb{Q}}} -e^{-i\varphi}\cos (\frac{\theta_{2} }{2}) {{\left| 1 \right\rangle }_{\mathbb{Q}}}}\ \ \ \ \ \ \ \ \ \ &{E^{e}_ {-}>E^{o}_ {-}} .
\end{array}} \right.
\end{equation}
For two quaternionic spinors $|\psi\rangle_{\mathbb{Q}}$ and $|\phi\rangle_{\mathbb{Q}}$  the scalar product is defined by
\begin{equation}\label{s-product}
\left\langle\phi|\psi \right\rangle_{\mathbb{Q}}:=\bar{\phi}^{\alpha}\psi_{\alpha}=\bar{\phi}_{1}\psi_{1}+\bar{\phi}_{2}\psi_{2},\quad\quad \psi_{1},\psi_{2},\phi_{1},\phi_{2}\in\mathbb{Q}.
\end{equation}
Note that right multiplication of quaternionic spinors with the nonzero quaternion $q$
yields the expression $\left\langle \phi q|\psi q \right\rangle_{\mathbb{Q}}=\bar{q}\left\langle\phi|\psi \right\rangle_{\mathbb{Q}}q$. The vector space $\mathcal{H}_{2}^{\mathbb{Q}}$ of quaternionic spinors with
this scalar product is a quaternionic Hilbert space. Let us consider the quaternionic spinor (\ref{quaterbit}) in useful form
\begin{equation}\label{quaterbit2}
|\psi \rangle_{\mathbb{Q}}=\left( {\begin{array}{*{20}{c}}
{{q_1}}\\
{{q_2}}
\end{array}} \right) = \frac{1}{{\sqrt {1 + {|x|^2}} }}\left( {\begin{array}{*{20}{c}}
1\\
x
\end{array}} \right)q,
\end{equation}
where $x=\frac{q_{0}\bar{q}_{1}}{|q_{1}|^2}$  come from stereographic projection $\mathcal{P}$ in Hopf fibration (\ref{hopf}) and $q$ is an unit quaternion ($q\bar{q}=1$). The distance between two  nonidentical, non orthogonal spinor states  $|\psi\rangle$ and $|\phi\rangle$ in base space $S^{4}$ is define as
\begin{equation}\label{distance}
\cos^{2}\frac{{{\Delta _{\phi \psi }}}}{2}=|\left\langle\phi|\psi \right\rangle_{\mathbb{Q}}|^{2}, \quad\quad 0 < \Delta  < \pi .
\end{equation}
In the   representation Eq. (\ref{quaterbit2}),  the Fubini-Study metric in quaternion spinor is define by \cite{Levay}
\begin{equation}\label{metric}
dl^{2}=g_{ij}dx^{i} \otimes dx^{j}=4(1-|\left\langle\psi+d\psi|\psi \right\rangle_{\mathbb{Q}}|^{2})=\frac{4d\bar{x}dx}{(1+|x|^{2})^{2}},
\end{equation}
and its corresponding connection is define by
\begin{equation}\label{connection}
\Gamma  =1-\left\langle\psi+d\psi|\psi \right\rangle_{\mathbb{Q}}=\bar{q}\left(\frac{Im(\bar{x}dx)}{1+|x|^{2}}\right)q+\bar{q}dq,
\end{equation}
where $Im(x)$ is imaginary part of quaternion $x$. The quantity $A=Im\frac{\bar{x}dx}{1+|x|^{2}}$ is a non-Abelian gauge field (one-form) which equivalent to the standard $SU(2)$ instanton with self-dual curvature and second Chern number $C_{2}=1$ and called Berry connection \cite{Levay}. Note that according to Eq. (\ref{connection}) the $\Gamma=0$ corresponding to parallel transformation of quaternionic phases. The differential equation  of the parallel transformation is determined with a suitable boundary condition in the equation (\ref{connection}). The standard path ordered solution for parallel transportation with initial and end points $q(0)=1$ and $q(\tau )$, respectively, is obtain by
\begin{equation}\label{parallel}
q(\tau)=\mathbf{P} \exp\left(-\int_{curve}{A}\right).
\end{equation}
For convenience in the next step, we consider the following state
\begin{equation}\label{quaterbit3}
|u\rangle_{\mathbb{Q}}=\left( {\begin{array}{*{20}{c}}
{{\cos(\frac{\theta}{2})}}\\
{{\sin(\frac{\theta}{2})p}}
\end{array}} \right)q,
\end{equation}
where $q,p $  are quaternionic phases with $|q|^2=|p|^2=1$ and the parametrization in therm of $\theta, p$ and $q$ has the same form as the well-known parametrization of a complex spinor associated with the Bloch sphere. According to Eq. (\ref{connection}) the connection for state Eq. (\ref{quaterbit3}) is given by
\begin{figure}
  \centering
  \includegraphics[width=16 cm]{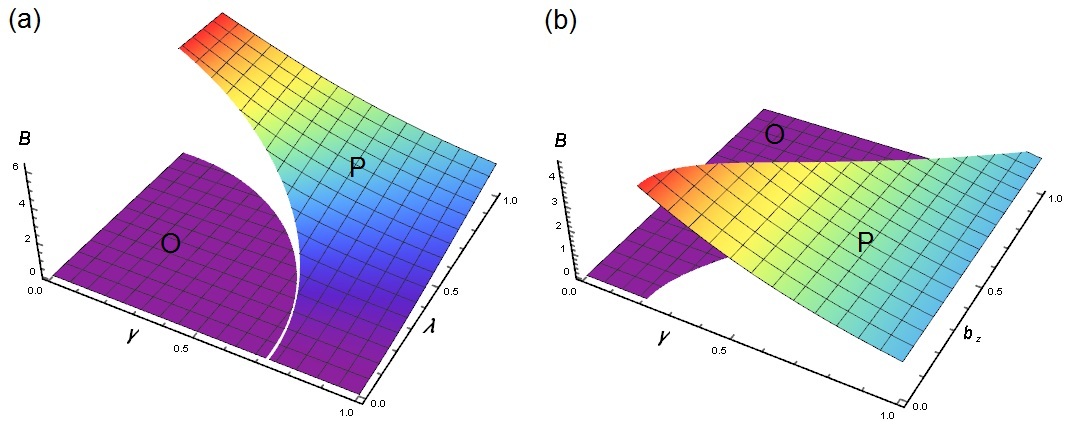}\\
 \caption{\small { (Color online.)  (a) Geometric phase of ground  state as a function $\gamma$ and $\lambda$ for Hamiltonian (\ref{hamiltonian}). The level crossing (white line) for parameters $J_{z}=0.15$, $D_{z}=0.2$, $b_{z}=0.1$.  (b) Geometric phase of ground  state as a function $\gamma$ and $b_z$ for Hamiltonian (\ref{hamiltonian}). The level crossing (white line) for parameters $J_{z}=0.1$, $D_{z}=0.4$, $\lambda=0.85$. }}\label{appendix Fig.3}
\end{figure}
\begin{equation}\label{cannection1}
\Gamma  =\frac{1-\cos(\theta)}{2}\bar{q}\left(Im(\bar{p}dp)\right)q+\bar{q}dq,
\end{equation}
and
\begin{equation}\label{bery}
A=\frac{1}{2}(1-\cos(\theta))Im(\bar{p}dp).
\end{equation}
The ground states of $H'$ in Eq. (\ref{qrstate}) takes the form of state (\ref{quaterbit3}), then the Berry connection of the ground states can be estimated as
\begin{equation}\label{berryconnection}
A = \left\{ {\begin{array}{*{20}{c}}
{(1 -\cos(\theta_{1} ))d\eta }&{E^{e}_ {-}<E^{o}_ {-}},\\
{0}&{E^{e}_ {-}>E^{o}_ {-}}.
\end{array}} \right.
\end{equation}
The parallel transformation  of ground states with respect to parameter $\eta$, gives the Berry phase $B=-\int A$ as follows
\begin{equation}\label{berryphase}
B = \left\{ {\begin{array}{*{20}{c}}
{ - 2\pi (1 - \cos(\theta_{1} )) }&{E^{e}_ {-}<E^{o}_ {-},}\\
{0}&{E^{e}_ {-}>E^{o}_ {-}.}
\end{array}} \right.
\end{equation}
\begin{figure}
  \centering
  \includegraphics[width=12 cm]{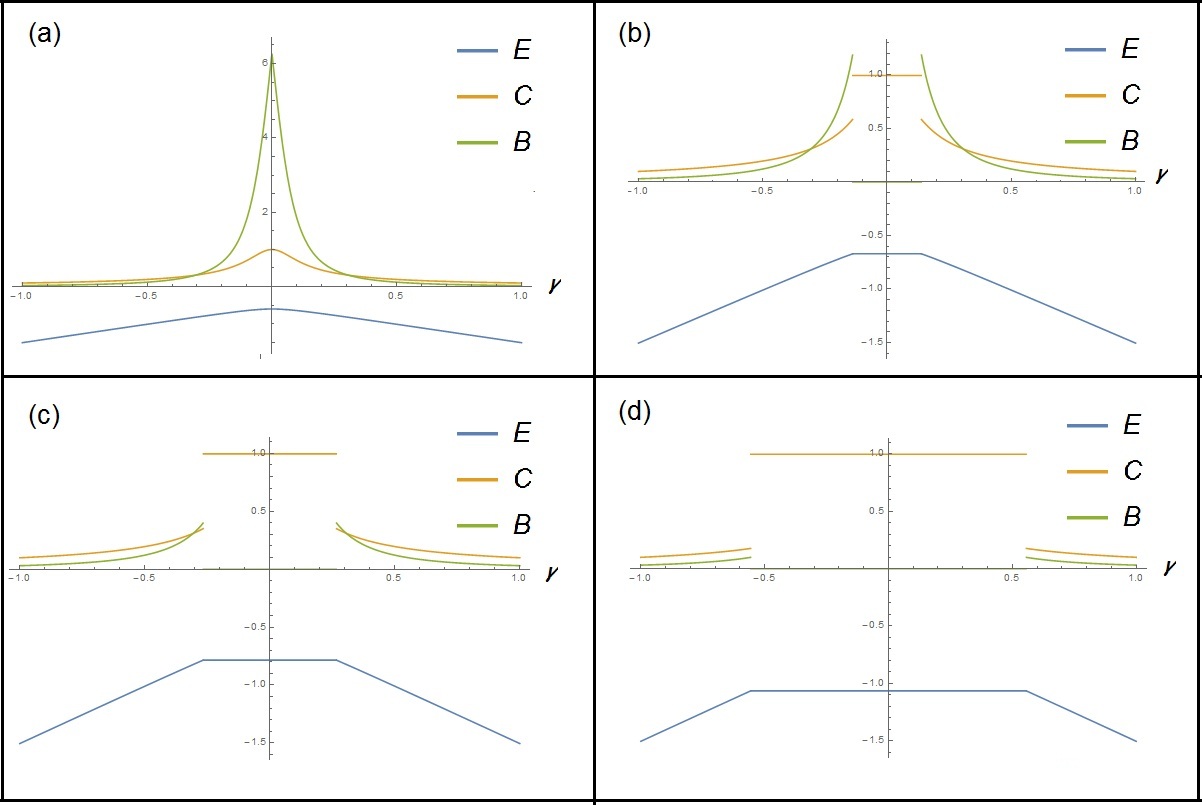}\\
 \caption{\small { (Color online.)  Energy, oncurrence and geometric phase of ground  state as a function of $\gamma$ for Hamiltonian (\ref{hamiltonian}) for parameters $J_{z}=0.5$, $\lambda=0.1$, $b_{z}=0.1$ and different value of $D_{z}$ (a) $D_{z}=0.4$, (b) $D_{z}=0.6$, (c) $D_{z}=0.8$ and (d) $D_{z}=1.2$ .}}\label{appendix Fig.4}
\end{figure}
\begin{figure}
  \centering
  \includegraphics[width=12 cm]{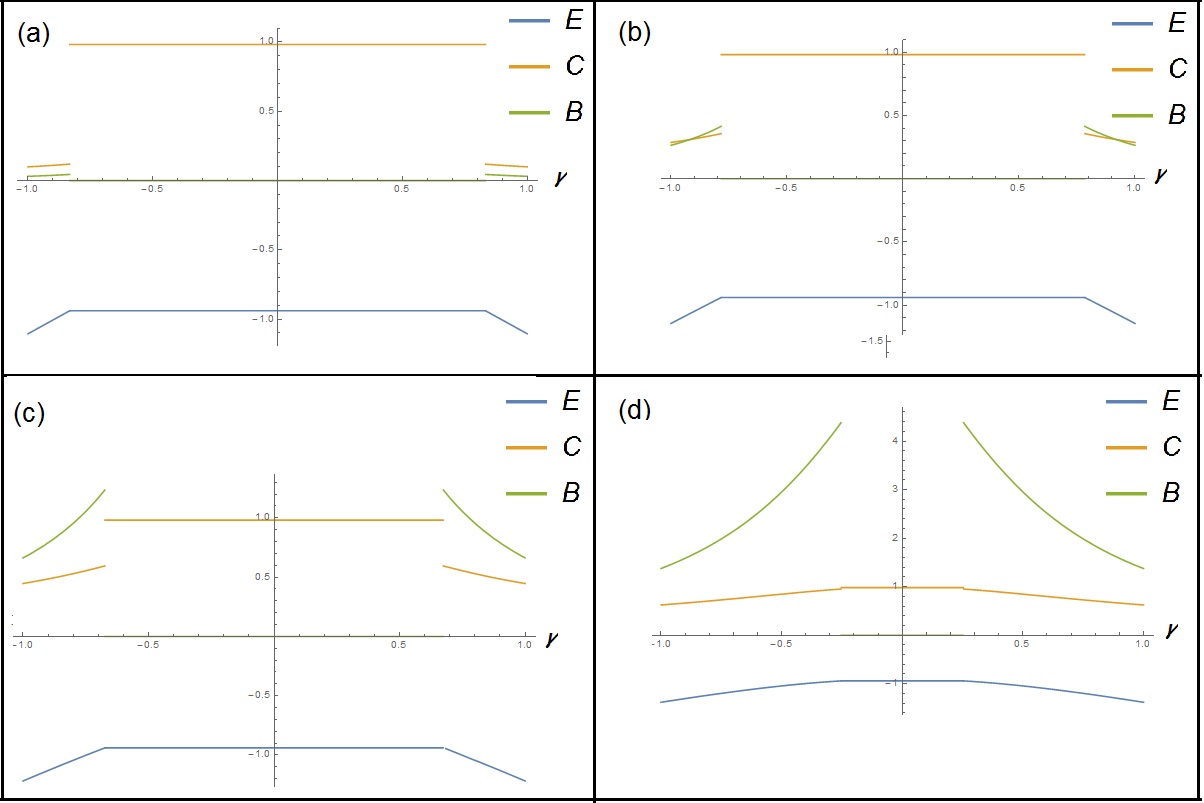}\\
 \caption{\small { (Color online.)  Energy, concurrence and geometric phase of ground  state as a function of $\gamma$ for Hamiltonian (\ref{hamiltonian}) for parameters $J_{z}=0.1$, $D_{z}=0.2$, $b_{z}=0.2$. and different value of $\lambda$. (a) $\lambda=0.1$, (b) $\lambda=0.3$, (c) $\lambda=0.6$ and (d) $\lambda=0.9$}}\label{appendix Fig.5}
\end{figure}
Fig. (\ref{appendix Fig.3}) shows the behavior of Berry phase for three regime. It seems that the Berry phase works as well an indicator to quantum phase transitions in Heisenberg Hamiltonian. However a comparison of the Berry phase and concurrence shows that in the regions that concurrence is not a appropriate indicator for the phase transition, the geometric phase is  a appropriate indicator, and vice versa, in the sense  that  in  regions where Berry phase does not show phase transition (i.e., the region $\lambda=0$ and $0<\gamma<1$ in Fig. (\ref{appendix Fig.3}-a) and the region $b_{z}=1$ and $0<\gamma<1$ in  Fig. (\ref{appendix Fig.3}-b)) the concurrence indicate the quantum  phase transition (see the corresponding region  in  Fig. (\ref{appendix Fig.2}-a) and   Fig. (\ref{appendix Fig.2}-b)). In other words, geometric phase and the ground state entanglement are complementary systems that can exhibit quantum phase transition.
Fig. (\ref{appendix Fig.4}) displays the ground state energy, concurrence and Berry phase as a function of anisotropic parameter for the nearest-neighbor spins in the Heisenberg  model (\ref{hamiltonian}) with different values of the DM interaction. It is clear that, for small value of $D_{z}$ the Berry phase have a significant changes in phase transition points, but the concurrence measure have a smooth  change in this region. It is interesting that  for a large value of $D_{z}$   concurrence changes is sharper than the Berry phase. Therefore, it is reasonable to conclude that  in small value of $D_{z}$ the Berry phase is good indicator than concurrence measure, and for large value of $D_{z}$ the concurrence measure  is good indicator than  Berry phase. On the other hand, by  increasing the DM interaction the phase transition occurs for large value of anisotropy parameter $\gamma$.
\par
In Fig. (\ref{appendix Fig.5}) we show the ground state energy, concurrence and Berry phase as a function of anisotropic parameter for the nearest-neighbor spins in the Heisenberg  model (\ref{hamiltonian}) with different values of the magnetic field $\lambda$.  It shows that for small value of magnetic field the concurrence measure is good indicator for quantum phase transition, but for large value of magnetic field the Berry phase show a sharp changes in phase transition point.

\section{Conclusion}
In summary, we have considered the anisotropic XYZ  Hamiltonian with  uniform and nonuniform external magnetic field and DM interaction. We saw that the geometric phase and concurrence measure are appropriate indicator for detecting the quantum phase transition in generalised Heisenberg model, but there are some phase transition regions that the geometric phase and concurrence measure do not change simultaneously even if the ground state changes according to level crossing points. This demonstrates that the geometric phase and concurrence measure individually do not capture a level crossing completely, which happens in ground state. In addition, we saw that the geometric phase and the ground state entanglement are complementary systems that can detect quantum phase transition. Moreover, we studied the geometric phase and concurrence measure in quaternionic representation that have geometric interpretation of this indicators. The geometric phase is proportional to parallel transportation in Hilbert space of two qubit states with Mannoury-Fubini-Study metric. Also, the concurrence measure is a quaternionic part of stereographic projection in quaternionic Hopf fibration.
We have showed that for $\lambda=0$ and $0<\gamma<1$ the entanglement is changed abruptly in phase transition point.  Therefor, the entanglement works as well an indicator in quantum phase transitions in this region. But for $\gamma=0$ and $0<\lambda<1$ the concurrence has not tangible change and it is not a suitable indicator for detect the phase transition.
 For small value of inhomogeneous magnetic field $b_{z}$ and $0<\gamma<1$ the Berry phase has intangible change, and for large value of $b_{z}$ the concurrence is changed abruptly in phase transition point.
We have showed  that the Berry phase can be used as an indicator to detect the quantum phase transitions in Heisenberg Hamiltonian, however a comparison of the Berry phase and concurrence shows that, where concurrence is not a good indicator for the phase transition, the geometric phase is  a appropirate one, and vice versa.
\par
We have showed  that, for small value of $D_{z}$ the Berry phase has a significant changes in phase transition points, but the concurrence measure have a smooth  change in this region. It is interesting that  for a large value of $D_{z}$ the concurrence has more significant chenges than Berry phase. Therefore, it is reasonable to conclude that  in small value of $D_{z}$ the Berry phase is more appropriate than concurrence measure in the sense of phase transition theory, and for large value of $D_{z}$ the concurrence measure  is suitable. On the other hand, by  increasing the DM interaction, the phase transition occurs for large value of anisotropy parameter $\gamma$. We plotted the ground state energy, concurrence and Berry phase as a function of anisotropic parameter for the nearest-neighbor spins in the Heisenberg model with different values of the magnetic field $\lambda$. 
Our  graphical results show that, for small values of magnetic field, the results of concurrence measure is more compatible than the results of ground state energy. On the other hand, for large values of magnetic field the Berry phase is more satisfied than the concurrence measure  in phase transition regions.

\par


\begin{thebibliography}{99}
\bibitem{Sachdev}
S. Sachdev: {\it{Quantum Phase Transitions}}, Cambridge Univ. Press, Cambridge, Second Edition. (2011)
\bibitem{Goldenfeld}
 N. Goldenfeld:  {\it{Lectures on phase transitions and the renormalization group}} . Urbana-Champaign, University of Illinois. (1992)
\bibitem{Mosseri}
 R. Mosseri, R. Dandoloff: Geometry of entangled states, Bloch spheres and Hopf fibrations. J. Phys. A Math. Gen. \textbf{34},  10243 (2001)
\bibitem{Bernevig}
B.A.  Bernevig, H.D. Chen: Geometry of the three-qubit state, entanglement and division algebras. J. Phys. A Math. Gen. \textbf{36}, 8325 (2003)
\bibitem{Oh}
S. Oh: Geometric phases and entanglement of two qubits with XY type interaction. Physics Letters A. \textbf{373},  644 (2009)
\bibitem{Berry}
 M.V. Berry: Quantal phase factors accompanying adiabatic changes. Proc. R. Soc. London Ser. A   \textbf{392},  45 (1984)
\bibitem{Shapere}
 A. Shapere, F. Wilczek:{\it{ Geometric Phases in Physics}}. World Scientific, Singapore.
(1989)
\bibitem{Najarbashi1}
G. Najarbashi, S. Ahadpour, M.A.Fasihi, Y. Tavakoli: Geometry of a two-qubit state and intertwining quaternionic conformal mapping under local unitary transformations. J. Phys. A Math. Theor.  \textbf{40}, 6481-6489 (2007)
\bibitem{Najarbashi2}
G. Najarbashi, B. Seifi, S. Mirzaei:Two- and three-qubit geometry, quaternionic and octonionic conformal maps, and intertwining stereographic projection. Quantum Inf. Process.  \textbf{15}, 509–528   (2016)
\bibitem{Najarbashi3}
G. Najarbashi, B. Seifi: Relation Between Stereographic Projection and Concurrence Measure in Bipartite Pure States. Int J Theor Phys.  10.1007/s10773-016-3071-2  (2016)
\bibitem{Levay}
P. L{\'e}vay: The geometry of entanglement: metrics, connections and the geometric phase. J. Phys. A  Math. Gen.  \textbf{37},  1821 (2004)
\bibitem{Oh1}
 S. Oh, Z. Huang, U. Peskin, S. Kais: Entanglement, Berry phases, and level crossings for the atomic Breit-Rabi Hamiltonian. Phys. Rev. A \textbf{78},  062106 (2008)
\bibitem{Bennett1}
C.H. Bennett, SJ. Wiesner: Communication via one-and two-particle operators on Einstein-Podolsky-Rosen states. Phys. Rev. Lett.  \textbf{69}, 2881 (1992)
\bibitem{Bennett2}
C.H. Bennett, G. Brassard, C. Cr\'epeau, R. Jozsa, A. Peres,  WK. Wootters:Teleporting an unknown quantum state via dual classical and Einstein-Podolsky-Rosen channels. Phys. Rev. Lett.   \textbf{70}, 1895 (1993)
\bibitem{Ekert}
A.K. Ekert: Quantum cryptography based on Bell’s theorem.Phys. Rev. Lett.  \textbf{ 67},  661 (1991)
\bibitem{Murao}
 M. Murao, D. Jonathan, M. B. Plenio, and V. Vedral: Quantum telecloning and multiparticle entanglement. Phys. Rev. A.  \textbf{ 59}, 156 (1999)
\bibitem{Schrodinger}
E. Schr{\"o}dinger: Probability relations between separated systems. Proc.
Camb. Phil. Soc.  \textbf{31}, 555 (1935)
\bibitem{Einstein}
A. Einstein,  B. Podolsky, N. Rosen : Can quantum-mechanical description of physical reality be considered complete?. Phys. Rev.  \textbf{47}, 777 (1935)
\bibitem{Bell}
J.S. Bell: On the Einstein-Podolsky-Rosen paradox. Physics \textbf{1}, 195 (1964)
\bibitem{Maleki}
Y. Maleki , F. Khashami, Y. Mousavi: Entanglement of Three-spin States in the Context of SU(2) Coherent States. Int J Theor Phys. \textbf{54}, 210 (2015)
\bibitem{Angelakis}
D. G. Angelakis, M. Christandl, A. Ekert, A. Kay, and S. Kulik: {\it{Quantum
Information Processing: From Theory to Experiment}}. Volume 199 NATO Science
Series: Computer and Systems Sciences, IOS Press. (2006)
\bibitem{Gunlycke}
D. Gunlycke, V. M. Kendon, V. Vedral, and S. Bose: Thermal concurrence mixing in a one-dimensional Ising model. Phys. Rev. A  \textbf{64}, 042302 (2001)
\bibitem{Yang}
Z. Yang, L. Yang, J. Dai, T. Xiang: Rigorous Solution of the Spin-1 Quantum Ising Model with Single-Ion Anisotropy. Phys. Rev. Lett.  \textbf{100}, 067203 (2008)
\bibitem{Lagmago}
G. L. Kamta. A. F. Starace:
Anisotropy and Magnetic Field Effects on the Entanglement of a Two Qubit Heisenberg XY Chain. Phys. Rev. Lett.  \textbf{88}, 107901 (2002)
\bibitem{Wang}
X. Wang:Thermal and ground-state entanglement in Heisenberg XX qubit rings. Phys. Rev. A  \textbf{66}, 034302 (2002)
\bibitem{Sun}
Y. Sun, Y. Chen, H. Chen: Thermal entanglement in the two-qubit Heisenberg XY model under a nonuniform external magnetic field. Phys. Rev. A  \textbf{68}, 044301 (2003)
\bibitem{Kao}
Z.C. Kao, J. Ng,  Y. Yeo: Three-qubit thermal entanglement via entanglement swapping on two-qubit Heisenberg XY chains. Phys. Rev. A  \textbf{72}, 062302 (2005)
\bibitem{Zhu}
S.L. Zhu: Scaling of Geometric Phases Close to the Quantum Phase Transition in the XY Spin Chain. Phys. Rev. Lett. \textbf{ 96}, 077206 (2006)
\bibitem{Asoudeh}
M. Asoudeh and V. Karimipour: Thermal entanglement of spins in an inhomogeneous magnetic field. Phys. Rev. A  \textbf{71}, 022308 (2005)
\bibitem{Zhang}
G.F. Zhang: Thermal entanglement and teleportation in a two-qubit Heisenberg chain with Dzyaloshinski-Moriya anisotropic antisymmetric interaction. Phys. Rev. A  \textbf{75}, 034304 (2007)
\bibitem{Zhang1}
G.F. Zhang, S.S. Li: Thermal entanglement in a two-qubit Heisenberg XXZ spin chain under an inhomogeneous magnetic field. Phys. Rev. A  \textbf{72}, 034302  (2005)
\bibitem{Kargarian}
M. Kargarian, R. Jafari, and A. Langari: Renormalization of entanglement in the anisotropic Heisenberg (XXZ) model. Phys. Rev. A  \textbf{77}, 032346  (2008)
\bibitem{Dzyaloshinsky}
I. Dzyaloshinsky: A thermodynamic theory of “weak” ferromagnetism of antiferromagnetics. J. Phys. Chem. Solids.  \textbf{4}, 241 (1958)
\bibitem{Moriya}
T. Moriya: New Mechanism of Anisotropic Superexchange Interaction. Phys. Rev. Lett.  \textbf{4}, 228 (1960)
\bibitem{Kheirandish}
F. Kheirandish, S.J. Akhtarshenas, H. Mohammadi: Effect of spin-orbit interaction on entanglement of two-qubit Heisenberg XYZ systems in an inhomogeneous magnetic field. Phys. Rev. A  \textbf{77}, 042309 (2008)
\bibitem{Wu}
L.A. Wu, D.A. Lidar: Universal quantum logic from Zeeman and anisotropic exchange  interactions. Phys. Rev. A  \textbf{ 66}, 062314 (2002)
\bibitem{Wu1}
L.A. Wu , D.A. Lidar: Dressed Qubits. Phys. Rev. Lett.  \textbf{91}, 097904  (2003)
\bibitem{Wootters}
 W. K. Wootters, Phys. Rev. Lett. \textbf{80}, 2245 (1998)
\bibitem{Hill}
S. Hill, W. K. Wootters, Phys. Rev. Lett. \textbf{78}, 5022  (1997)
\bibitem{Vedral}
V. Vedral : {\it{Modern Foundations of  Quantum optics}}, University of Leeds, UK
,Imperial College Press
. (2005)
\end{thebibliography}
\end{document}